\documentclass[,final]{aipproc}

\layoutstyle{6x9}

\def\ignore#1{{}}
\newcommand{\bea}{\begin{eqnarray}}
\newcommand{\eea}{\end{eqnarray}}

\begin{document}

\title{Operators in ultraviolet completions 
for \\ Electroweak collective symmetry breaking
}

\classification{12.60.Cn \hfill
OU-HET 641/2009}

\author{Nobuhiro Uekusa}{
  address={Department of Physics, 
Osaka University,
Toyonaka, Osaka 560-0043
Japan}
}

\begin{abstract}
In a gauge theory with SU(5) broken to
SU(2)$\times$U(1),
the Higgs mass squared receives only logarithmic
divergence from all scalar-gauge interactions
at one loop.
The same pattern of gauge symmetry breaking
can be achieved without any scalar fields
in a five-dimensional model.

\end{abstract}

\maketitle


Quadratic divergence is one of the fundamental problems
of the standard model of particle physics.
The structure of quantum loop corrections is different
between scalars and fermions.
It is chiral symmetry that protects fermions from 
having quadratic divergence.
Scalar fields do not have such a symmetry in the standard model. 
Applying symmetry principle to scalar fields has been
a clue to understand physics beyond the standard model. 
If scalar fields were Nambu-Goldstone bosons in broken global 
symmetry, they are exactly massless and
it is led to little Higgs scenarios~%
\cite{ArkaniHamed:2001nc,ArkaniHamed:2002qy}.

In little Higgs scenarios, if gauge and Yukawa
couplings are vanishing,
Higgs fields have derivative couplings
and they have shift symmetry so that
there is no potential.
The scalar fields are described in a non-linear sigma
model.
The assumption that the global symmetry is
explicitly broken only when two or more couplings
are non-vanishing leads to a potential
with at most logarithmic divergence.
This collective symmetry breaking mechanism has been 
studied in many papers~%
\cite{Kaplan:2003uc,Cheng:2003ju,
Low:2004xc,Schmaltz:2005ky}.
In the littlest Higgs model, the resulting operator of
scalar-gauge interactions is, for example,
\bea
   W_{1\mu} W_2^{\,\mu} h^\dag h ,
   \label{basic}
\eea
where $h$ is the Higgs field and
$W_{1\mu}$ and $W_{2\mu}$ are $[\textrm{SU}(2)]^2$ gauge bosons.
At one loop, the gauge bosons 
do not produce the quadratic divergence for
the Higgs mass squared through the interaction (\ref{basic}).
While the collective symmetry breaking mechanism requires two or
more couplings,
such a group as $[\textrm{SU}(2)]^2$ could be
a subgroup of a single group.
It should be clarified whether
operators such as Eq.~(\ref{basic}) can be derived
in a renormalizable gauge theory with a single large group broken to 
two or more subgroups.

Recently, in Ref.~\cite{Csaki:2008se}
a weakly-coupled renormalizable ultraviolet completion of a little 
Higgs scenario was proposed.
It was claimed that heavy modes are integrated out
and that the remaining theory is a non-linear sigma model due to the 
Coleman-Wess-Zumino theorem~\cite{Coleman:1969sm,Callan:1969sn}.
In the model,
the Higgs mass squared receives radiative correction 
of the order of 100~GeV and the non-linear sigma model
has the decay constant $f\sim 1$~TeV and 
the ultraviolet momentum cutoff $\Lambda\sim 10$~TeV.
Also, in an explicit example, several 
effective couplings with the form (\ref{basic}) were shown.
To pursue this possibility of high energy theory for 
no quadratic divergence as an extension of the collective 
symmetry breaking mechanism,
it would be important to examine more couplings.

In the following, we give all the couplings of
scalar-gauge interactions in a weakly-coupled gauge theory
beyond the ultraviolet momentum cutoff
of models where the collective symmetry breaking
mechanism protects the Higgs mass squared against
having quadratic divergence.
As another aspect, we study a possibility of gauge symmetry breaking 
by boundary conditions 
without any scalar fields.

\vspace{4ex}

We start with a renormalizable gauge theory with a single group
which is broken by the vacuum expectation values
of scalar fields, aiming for having
scalar-gauge interactions of the form
\bea
   g^2 W_{1\mu} W_2^{\,\mu} h^\dag h ,
   \label{basic2}
\eea
where a gauge coupling is denoted as $g$.
For the interaction (\ref{basic2}), 
the one-loop diagram for
the Higgs mass squared has propagators of
two gauge bosons.
Therefore
the divergence is at most logarithmic.
Since Eq.~(\ref{basic2}) is written as
\bea 
    g^2 \left({W_{1\mu}+W_{2\mu}\over \sqrt{2}}\right)^2 h^\dag h
  -g^2 \left({W_{1\mu}-W_{2\mu}\over \sqrt{2}}\right)^2 h^\dag h ,
   \label{basic3}
\eea
the absence of the quadratic divergence for the Higgs mass
squared is also regarded as a cancellation.
In little Higgs models, the collective symmetry breaking
mechanism can produce the operator (\ref{basic2}).
In such a model, the self-interaction of the Higgs fields obeys
a non-linear sigma model.
The Lagrangian terms of expansion at lower order are
$|\partial_\mu h|^2
+f^{-2}|\partial_\mu h|^2 h^\dag h$.  
The quadratic divergence of one-loop contribution to
the kinetic term is 
$\Lambda^2/(16\pi^2f^2)$,
where $\Lambda$ is the ultraviolet momentum cutoff.
For the non-linear sigma model to be viable,
the cutoff has the upper bound
$\Lambda < 4\pi f$.
The dominant correction to the Higgs mass squared
is 
$(g^4/(16\pi^2)) f^2 \log 
(\Lambda^2/f^2)$.
If the correction is of the order of 100~GeV,
the decay constant is obtained as
$f\sim 1$~TeV and then $\Lambda\sim 10$~TeV.
Thus our starting theory is defined at higher
scales than 10~TeV.

We choose the single gauge group as SU(5).
The gauge symmetry is broken by the vacuum expectation
values of two types of scalar fields.
We assume that these two sectors 
yield separate gauge symmetry breakings
\bea
\textrm{SU}(5)\to [\textrm{SU(2)}\times\textrm{U(1)}]^2 ,
\qquad
\textrm{SU(5)$\to$SO(5)} .
\eea 
The Higgs fields are included in the sector of the
breaking SU(5)$\to$SO(5).
The scalar fields in the other sector are decoupled
to the Higgs fields.
The scalar fields we discuss are only in the 
sector of SU(5)$\to$SO(5).

The scalar field responsible for 
the breaking SU(5)$\to$SO(5)
is 
a scalar \underline{15} which transforms
as a complex symmetric matrix 
$S \to S'=U S U^T$
under SU(5).
The potential is written as
$V = -M^2 \textrm{Tr} \left[SS^\dag\right]
     +\lambda_1 (\textrm{Tr}\left[SS^\dag\right])^2
     +\lambda_2 \textrm{Tr}[(SS^\dag)^2]$,
where $M$, $\lambda_1$ and $\lambda_2$ are the coupling constants.
From the stationary condition,
the nonzero expectation value is given by
$SS^\dag =  f_S^2 \textrm{\bf 1}_5$ (for $10\lambda_1 + 2\lambda_2 \neq 0$),
which is proportional to the identity 
matrix. This means symmetry breaking to O(5).
Here the decay constant is defined as
$f_S^2 \equiv M^2/(10\lambda_1 +2\lambda_2)$.
For the vacuum expectation value
\bea
   \langle S\rangle =
     f_S \left(\begin{array}{ccc}
       &&  {\bf 1}_2 \\
       &1 &  \\
       {\bf 1}_2 && \\
     \end{array}\right) ,
      \label{vev}
\eea
the global SU(5)$\times$U(1) is broken to SO(5).
Fluctuations around the vacuum expectation value (\ref{vev}) are parameterized as
$S=\langle S\rangle + \bar{S}$ with 
\bea
 \bar{S} = iN + R , \qquad
   N=\left(\begin{array}{ccc}
      \phi & h & \chi \\ 
      h^T & K_i & h^\dag \\ 
      \chi^T & h^* & \phi^\dag \\
     \end{array}\right) , 
 \quad
 R =\left(\begin{array}{ccc}
    \Phi & H & X \\ 
    H^T & K_r & H^\dag \\ 
    X^T & H^* & \Phi^\dag \\ 
    \end{array} \right) .
\eea
which are composed of the two 2$\times$2 complex symmetric tensors
$\phi$ and $\Phi$, the two complex doublet fields $h$ and $H$,
the two 2$\times$2 Hermite tensors $\chi$ and $X$ and
the two real fields $K_i$ and $K_r$.
The fields in the matrix $R$
have large masses $\sim M$.

After integrating out heavy fields, 
we obtain the Lagrangian term of effective vertex 
of gauge-scalar couplings
$V_{\textrm{\tiny eff}}$ as~%
\cite{Uekusa:2008ag}
\bea
 -V_{\textrm{\tiny eff}}
  &\!\!\! =\!\!\!&
      (\textrm{${1\over 4}$}g^2 W_1^a W_2^a 
 +\textrm{${1\over 8}$}g'{}^2 B_1 B_2) h^\dag h
 +\textrm{${1\over 4}$}
      g^2 W_1^a W_2^a ( \textrm{Tr}\left[\phi\phi^\dag\right]
        +\textrm{Tr}\left[\chi^2\right] )
\nonumber
\\
&& +\textrm{${1\over 400}$} g'{}^2 
  (B_1^2 + B_2^2 + 49 B_1 B_2 )\textrm{Tr} \left[\phi\phi^\dag\right]
   -\textrm{${3\over 50}$}
    g'{}^2 (B_1-B_2)^2 \textrm{Tr} (\chi^2) .
    \label{main1}
\eea
The Higgs field $h$ has gauge couplings
of the form given in Eq.~(\ref{basic2}).
Thus for the mass of the Higgs field, 
there is no quadratic divergence from gauge boson loop.
On the other hand, $\textrm{Tr}(\phi\phi^\dag)$, $\textrm{Tr}(\chi^2)$
have interacting terms with $B_1^2$ and $B_2^2$. Due to these U(1) factors, the masses of $\phi$ and $\chi$ receive quadratic divergence.

\vspace{4ex}

As another aspect,
we study the case in which 
the two breaking sectors in higher energy scales are spatially separated.
At $y=0$, SU(5) is broken to 
$[\textrm{SU(2)}\times\textrm{U(1)}]^2$ and
at $y=L$, SU(5) is broken to SO(5).
Here $y$ is the coordinate
of the fifth dimension and the fundamental region
is $0\leq y \leq L$.
We assume that the five-dimensional spacetime is flat.
It has been shown that  at a single boundary
SU(5) cannot be broken
to $\left[ \textrm{SU}(2)\times \textrm{U}(1)\right]$
by boundary conditions
unless additional scalar fields are introduced~%
\cite{Sakai:2006qi}. 
It is needed to examine the consistency of 
possible boundary conditions to yield gauge symmetry breaking
$\textrm{SU}(5)\to \left[ \textrm{SU}(2)\times \textrm{U}(1)\right]^2$
at $y=0$ and
$\textrm{SU}(5)\to \textrm{SO}(5)$ at $y=L$. 
In order to realize the symmetry breaking above,
we assign Neumann condition for the gauge bosons of the generators 
$T_1$, $T_2$, $T_3$, $T_8$, $T_{15}$, $T_{22}$,
$T_{23}$, $T_{24}$ at $y=0$ and  
$T_{\bar{1}},\cdots, T_{\bar{10}}$ given by
$T_{\overline{1}}=\textrm{${1\over \sqrt{2}}$}
(T_1 -T_{22})$,
$T_{\overline{2}}=\textrm{${1\over \sqrt{2}}$}
   (T_2 +T_{23})$, 
$T_{\overline{3}}=\textrm{${1\over \sqrt{2}}$}
  (T_4 -T_{13})$,
$T_{\overline{4}}=\textrm{${1\over \sqrt{2}}$}
  (T_5 -T_{14})$,   
$T_{\overline{5}}=\textrm{${1\over \sqrt{2}}$}
  (T_6 -T_{20})$,
$T_{\overline{6}}=\textrm{${1\over \sqrt{2}}$}
  (T_7 -T_{21})$,
$T_{\overline{7}}=\textrm{${1\over \sqrt{2}}$}
  (T_{11} -T_{16})$,
$T_{\overline{8}}=\textrm{${1\over \sqrt{2}}$}
  (T_{12} -T_{17})$,
$T_{\overline{9}}
 = \textrm{${1\over \sqrt{2}}$}(T_3 +\textrm{${\sqrt{6}\over 4}$}T_{15}
      -\textrm{${\sqrt{10}\over 4}$}T_{24})$,
$T_{\overline{10}}
    = \textrm{${1\over \sqrt{2}}$}(\textrm{${\sqrt{3}\over 3}$}
   T_8 +\textrm{${5\sqrt{6}\over 12}$}T_{15}
      +\textrm{${\sqrt{10}\over 4}$}T_{24})$,  
at $y=L$ and Dirichlet condition for the other gauge bosons.
Only the fields with Neumann condition at both boundaries have zero modes.
The generators of zero modes are
$T_{\overline{1}} , T_{\overline{2}},
  T_{\overline{9}}, T_{\overline{10}}$.
These generators form SU(2)$\times$U(1) algebra.
The gauge transformation laws are written as
$\delta A_M^{a} =\partial_M \epsilon^{a}
    + gf^{abc} A_M^{b} \epsilon^{c}
  + gf^{a\hat{b}\hat{c}} A_M^{\hat{b}} 
  \epsilon^{\hat{c}}$,
$\delta A_M^{\hat{a}} =
  \partial_M \epsilon^{\hat{a}}
    + gf^{\hat{a}b\hat{c}} A_M^{b}\epsilon^{\hat{c}}
  + gf^{\hat{a}b\hat{c}} A_M^{b} 
  \epsilon^{\hat{c}}$,
where $a$ and $\hat{a}$ indicate the generators of the subgroup and
the coset, respectively.
At $y=0$, $a$ and $\hat{a}$ represent the indices for
$\left[ \textrm{SU}(2)\times \textrm{U}(1)\right]^2$
and $\textrm{SU}(5)/\left[ \textrm{SU}(2)\times \textrm{U}(1)\right]^2$, respectively.
At $y=L$, $a$ and $\hat{a}$ represent the indices for
SO(5) and SU(5)/SO(5), respectively.
The boundary conditions at $y=0$ and $y=L$ are collectively written 
with the $a$ and $\hat{a}$. 
The left- and right-hand sides in the transformation laws have
the boundary conditions shown in Table~\ref{bct55}.
\begin{table}[htb]
\caption{The boundary conditions for 
$\textrm{SU}(5)\to\textrm{SU}(2)\times\textrm{U}(1)$.} \label{bct55}
\begin{tabular}{c|c|c}
\hline
 & LHS & RHS \\ \hline
$A_\mu^a$ & N & N $+$ NN $+$ DD \\
$A_y^a$ & D & D $+$ DN $+$ ND \\ \hline
\end{tabular} ~~
\begin{tabular}{c|c|c}
\hline
 & LHS & RHS \\ \hline
$A_\mu^{\hat{a}}$ & D & D $+$ DN $+$ ND \\
$A_y^{\hat{a}}$ & N & N $+$ NN $+$ DD \\ \hline
\end{tabular}
\end{table}
From Table~\ref{bct55}, it is found that all the boundary conditions are 
consistent with the gauge transformations.
The result of the consistency in this case is also seen from the fact that 
Neumann and Dirichlet conditions imposed at each boundary 
can be assigned by
the automorphism of orbifolds.

\vspace{4ex}

We have derived all the couplings
of scalar-gauge interactions
relevant to scalar mass corrections
 via gauge boson loop in high energy models.
We have found that while the Higgs fields are protected
with the coupling of the form (\ref{basic}) 
from having quadratic 
divergence, the other fields receive quadratic divergence through
loop of the U(1) gauge boson.
We have also considered a possibility of gauge symmetry breaking 
by boundary conditions. 
In our assignment for boundary conditions,
the same gauge symmetry breaking as in the vacuum expectation value
is produced in a consistent way with local gauge transformations.


\begin{theacknowledgments}
 This work is supported by Scientific Grants 
from the Ministry of Education
and Science, Grant No.~20244028.
\end{theacknowledgments}



\bibliographystyle{aipproc}   

\bibliography{sample}

\IfFileExists{\jobname.bbl}{}
 {\typeout{}
  \typeout{******************************************}
  \typeout{** Please run "bibtex \jobname" to optain}
  \typeout{** the bibliography and then re-run LaTeX}
  \typeout{** twice to fix the references!}
  \typeout{******************************************}
  \typeout{}
 }



\end{document}